\documentclass[aps,prc,showpacs,superscriptaddress,showkeys,twocolumn]{revtex4}
\usepackage{latexsym}
\usepackage{amssymb}
\usepackage{amsmath}
\usepackage{graphicx}
\usepackage{bm}
\usepackage{epsf}
\usepackage{epstopdf}

\begin{document}

\title{Nuclear pasta in hot and dense matter and its influence on
the equation of state for astrophysical simulations}

\author{Fan Ji}
\affiliation{School of Physics, Nankai University, Tianjin 300071, China}
\author{Jinniu Hu}~\email{hujinniu@nankai.edu.cn}
\affiliation{School of Physics, Nankai University, Tianjin 300071, China}
\affiliation{Strangeness Nuclear Physics Laboratory, RIKEN Nishina Center, Wako, 351-0198, Japan}
\author{Shishao Bao}~\email{bao_shishao@163.com}
\affiliation{School of Physics and Information Engineering, Shanxi Normal University, Linfen 041004, China}
\author{Hong Shen}~\email{shennankai@gmail.com}
\affiliation{School of Physics, Nankai University, Tianjin 300071, China}

\begin{abstract}
We explore the properties of nuclear pasta appearing in supernova matter, i.e.,
matter at finite temperature with a fixed proton fraction.
The pasta phases with a series of geometric shapes are studied using the compressible
liquid-drop (CLD) model, where nuclear matter separates into a dense liquid phase of nucleons
and a dilute gas phase of nucleons and $\alpha$ particles.
The equilibrium conditions for two coexisting phases are derived by
minimization of the total free energy including the surface and Coulomb
contributions, which are clearly different from the Gibbs conditions for
phase equilibrium due to the finite-size effects.
Compared to the results considering only spherical nuclei, the inclusion of pasta
phases can delay the transition to uniform matter and enlarge the region of
nonuniform matter in the phase diagram. The thermodynamic quantities obtained
in the present calculation with the CLD model are consistent with those in
the realistic equation of state table for astrophysical simulations using
the Thomas--Fermi approximation.
It is found that the density ranges of various pasta shapes depend on
both the temperature $T$ and the proton fraction $Y_p$.
Furthermore, the nuclear symmetry energy and its density dependence may play crucial
roles in determining the properties of pasta phases.
Our results suggest that the pasta phase diagram is most sensitively
dependent on the symmetry energy slope $L$ especially in the low-$Y_p$
and high-$T$ region.
\end{abstract}

\maketitle


\section{Introduction}
\label{sec:1}

Core-collapse supernovas are one of the most fascinating phenomena in the
universe, and lead to the formation of neutron stars or black holes.
During the past decades, great efforts have been devoted to numerical
simulations of gravitational collapse of massive stars~\cite{Burr13,Jank12,Jank16},
in which the equation of state (EOS) of dense matter is an essential ingredient.
The EOS plays an important role in understanding the dynamics of supernova explosions,
which requires information over very wide ranges of temperature, proton fraction,
and baryon density (see, e.g., Table 1 of Ref.~\cite{Shen11}).
In the full thermodynamic parameter space, the nuclear matter exhibits a rich and
complex phase diagram. At low temperatures and subsaturation densities,
the matter is nonuniform where heavy nuclei are formed to lower the free energy
of the system. When the density is beyond $\approx 1/2$ nuclear saturation density,
heavy nuclear clusters tend to dissolve into a uniform nucleon liquid.
It is likely that nonspherical nuclei, known as pasta phases, may appear as the
density approaches the phase transition to uniform
matter~\citep{Pais12,Avan10,Wata09,Okam13,Bao15}.
However, heavy nuclei cannot be formed above a critical temperature,
where the matter is a mixture of free nucleons and light clusters together with
leptons~\citep{Capl16,Schu13}.
At densities much higher than nuclear saturation density, non-nucleonic degrees of
freedom like hyperons and quarks may occur and soften the EOS of dense matter~\citep{Oert17}.

It is a challenge to construct a realistic EOS covering the whole range of thermodynamic
conditions for numerical simulations of core-collapse supernovas.
Currently, there are various EOSs available for astrophysical simulations
such as core-collapse supernovas and neutron-star mergers 
(see, e.g., Ref.~\citep{Oert17} for a recent review).
One of the most commonly used EOSs is the Lattimer-Swesty EOS~\citep{Latt91},
which employed a compressible liquid-drop (CLD) model with Skyrme forces to describe
heavy nuclei in nonuniform matter. Recently, the approach of Lattimer and Swesty
was extended and improved by Schneider \emph{et al}.~\citep{Schn17,Schn19} for computing
many EOSs based on Skyrme-type parametrizations of the nuclear forces.
Another commonly used EOS is often referred to as the Shen EOS~\citep{Shen98a,Shen98b,Shen11},
which was based on the relativistic mean-field (RMF) model and Thomas--Fermi approximation
with a parametrized nucleon distribution for the description of nonuniform matter.
A similar Thomas--Fermi approximation with realistic nuclear forces was used to construct
the EOS table by Togashi \emph{et al}.~\citep{Toga17} recently.
In these realistic EOSs for astrophysical simulations, the nonuniform matter
at intermediate densities is treated using the single nucleus
approximation (SNA)~\citep{Burr84}.
There are also several EOSs that were developed beyond the SNA by
including multiple nuclei in the framework of nuclear statistical equilibrium (NSE)~\citep{Hemp10,Furu11,Furu13,Furu17a,Stei13}.
Usually only spherical heavy nuclei are considered in constructing the EOS
tables. In the present work, we intend to explore the influence of nonspherical
pasta phases on the EOS for astrophysical simulations.

The appearance of nuclear pasta is mainly caused by the competition between
the surface and Coulomb energies of heavy nuclei. As a result, the stable nuclear
shape in nonuniform matter may change from droplet to rod, slab, tube, and
bubble with increasing baryon density. Nuclear pasta phases are expected to
occur both in core-collapse supernova matter with fixed proton fraction
at finite temperature and in the inner crust of neutron stars where
neutron-rich matter is in $\beta$ equilibrium at zero temperature.
Over the past decades, the properties of pasta phases have been studied
using various methods, such as the liquid-drop model~\cite{Rave83,Hash84,Wata00,Bao14a}
and the Thomas--Fermi approximation~\citep{Avan10,Oyam07,Gril12,Pais15}.
Generally, the Wigner--Seitz approximation with typical geometric shapes of
nuclear pasta is employed to simplify the calculations. For more realistic
description, there are some studies that have not explicitly
assumed any geometric shape and performed fully three-dimensional calculations
for nuclear pasta based on the Thomas--Fermi approximation~\cite{Okam13,Will85,Okam12},
Hartree-Fock approach~\citep{Pais12,Magi02,Newt09,Schu13,Sage16,Fatt17}, and
molecular dynamics method~\citep{Wata09,Capl17,Maru98,Sono08,Schn13,Schn14}.
It is noteworthy that nuclear symmetry energy and its slope could significantly
affect the pasta phase structure and crust-core transition of neutron
stars~\citep{Bao15,Oyam07,Gril12}.

For the pasta phases in supernova matter, Pais \emph{et al}.~\citep{Pais15}
performed calculations and compared results using three different methods:
the coexisting phases (CP) method, the CLD model,
and the self-consistent Thomas--Fermi approximation.
The CP method is relatively simple, whereby two coexisting phases satisfy the
Gibbs conditions for phase equilibrium, whereas the surface and Coulomb
contributions are perturbatively taken into account~\citep{Avan10,Avan12}.
In the CLD model, the surface and Coulomb contributions are treated
in a more consistent manner, and are included in the minimization procedure
and lead to some additional terms in the equilibrium conditions~\citep{Bao14b}.
The Thomas--Fermi approximation describes the nucleon distributions of pasta phases
in a realistic way, whereby the finite-size effects are treated self-consistently.
Recently, the impact of nuclear pasta on the neutrino scattering rates has been discussed
for core-collapse supernovas and protoneutron star evolution~\citep{Rogg18,Horo16},
and it was found that the presence of nuclear pasta could alter the late-time neutrino
signal from supernovas.
The elastic properties of nuclear pasta are presently interesting to some researchers
for their relevance to gravitational wave searches both from supernova and neutron-star
mergers, which motivates calculations of the pasta phase
diagram~\citep{Abbo17b,Peth20,Peth19,Capl18}.
Therefore, it is interesting and important to investigate under
which conditions the pasta phases can occur.

In this article, we have two aims. The first is to investigate the properties
of pasta phases that occur in supernova matter, while the effects of nuclear symmetry
energy are examined by using two RMF models, namely, the TM1 and TM1e
parametrizations~\citep{TM1,Bao14b},
which have the same properties of symmetric nuclear matter
but different behaviors of the symmetry energy.
The second is to explore the influence of nuclear pasta on the EOS for astrophysical
simulations. We perform calculations of nonuniform matter using the CLD model,
where a nuclear liquid coexists with a dilute gas consisting of free nucleons
and $\alpha$ particles employing a sharp interface.
By comparing the results with and without pasta phases, we analyze the possible
impact from nuclear pasta on the phase diagram and thermodynamic quantities.
Since both the TM1 and TM1e models have been employed in constructing
the EOS tables for core-collapse supernova simulations using a parametrized
Thomas--Fermi approximation~\citep{Shen11,Shen20}, it is possible to examine
the difference between the present results using the CLD method and the values
from realistic EOS tables, so that the uncertainty due to different descriptions
of nonuniform matter can be estimated quantitatively.

It is necessary to check the nuclear model by recent developments in
astrophysical observations.
One strong constraint coming from the mass measurements of massive
pulsars~\citep{Demo10,Fons16,Anto13,Crom19} requires the maximum
neutron-star mass to be larger than $\approx 2 M_\odot$.
We notice that the TM1 and TM1e models predict maximum neutron-star
masses of $2.18 M_\odot$ and $2.12 M_\odot$, respectively.
Recently, the first detection of gravitational waves from a binary neutron-star merger,
known as GW170817, provided valuable constraints on the tidal deformability~\citep{Abbo17,Abbo18},
which also restricts the radius of a canonical $1.4 M_\odot$ neutron star
as $R_{1.4}<13.8$ km~\cite{Tews18,Zhu18,De18,Fatt18,Mali18}.
More recently, the second detection of gravitational waves, GW190425, was
reported by the LIGO and Virgo Collaborations~\citep{Abbo19}.
The latest observations by the Neutron Star Interior Composition Explorer (NICER)
for PSR J0030+0451 provided a simultaneous measurement of the mass and radius of
a neutron star~\citep{Rile19,Mill19}.
It is interesting to notice that constraints on the neutron-star radius
from various observations are consistent with each other.
In our previous work~\citep{Ji19}, we studied the correlation between the neutron-star
radius and the slope parameter $L$ of symmetry energy using a family of
RMF models generated from the TM1 parametrization.
The TM1e model with $L=40$ MeV predicts a radius
of $R_{1.4}=13.1$ km that is well within the current constraints,
whereas the original TM1 model with $L=110.8$ MeV results in a much larger
radius of $R_{1.4}=14.2$ km. Furthermore, the neutron-star maximum mass and
tidal deformability predicted by the TM1e model are also compatible
with observational constraints.
In the present study, we employ the TM1e model with $L=40$ MeV
to perform calculations of nonuniform matter including pasta phases,
whereas the results from the original TM1 model with $L=110.8$ are also
presented to examine the influence of the density dependence of symmetry energy.

This article is organized as follows. In Sec.~\ref{sec:2},
we briefly review the RMF model used and describe the CLD method
for the description of pasta phases in hot and dense matter.
In Sec.~\ref{sec:3}, the results of nuclear pasta and its influence
on the EOS are discussed.
Finally, the conclusions are presented in Sec.~\ref{sec:4}.

\section{Formalism}
\label{sec:2}

We study the nuclear pasta phases at finite temperature based on the CLD method,
where the RMF model with extended TM1 parametrization is used for the nuclear
interaction~\citep{Bao14b}. In the RMF approach, nucleons interact via the exchange
of various mesons including the isoscalar-scalar meson $\sigma$,
isoscalar-vector meson $\omega$, and isovector-vector meson $\rho$.
The nucleonic Lagrangian density reads
\begin{eqnarray}
\label{eq:LRMF}
\mathcal{L} & = & \sum_{i=p,n}\bar{\psi}_i \left[
i\gamma_{\mu}\partial^{\mu}-\left(M+g_{\sigma}\sigma\right)
 \right. \notag \\  & & \left.
-\gamma_{\mu} \left(g_{\omega}\omega^{\mu}
+\frac{g_{\rho}}{2} \tau_a\rho^{a\mu}\right) \right]\psi_i  \notag \\
&& +\frac{1}{2}\partial_{\mu}\sigma\partial^{\mu}\sigma -\frac{1}{2}%
m^2_{\sigma}\sigma^2-\frac{1}{3}g_{2}\sigma^{3} -\frac{1}{4}g_{3}\sigma^{4}
\notag \\
&& -\frac{1}{4}W_{\mu\nu}W^{\mu\nu} +\frac{1}{2}m^2_{\omega}\omega_{\mu}%
\omega^{\mu} +\frac{1}{4}c_{3}\left(\omega_{\mu}\omega^{\mu}\right)^2  \notag
\\
&& -\frac{1}{4}R^a_{\mu\nu}R^{a\mu\nu} +\frac{1}{2}m^2_{\rho}\rho^a_{\mu}\rho^{a\mu}
 \notag \\  & &
+\Lambda_{\mathrm{v}} \left(g_{\omega}^2
\omega_{\mu}\omega^{\mu}\right)
\left(g_{\rho}^2\rho^a_{\mu}\rho^{a\mu}\right),
\end{eqnarray}
where $W^{\mu\nu}$ and $R^{a\mu\nu}$ denote the antisymmetric field tensors for
$\omega^{\mu}$ and $\rho^{a\mu}$, respectively.
Under the mean-field approximation, the meson fields are treated as classical
fields and the field operators are replaced by their expectation values.
In a static system, the nonvanishing expectation values of meson fields are
$\sigma =\left\langle \sigma \right\rangle$,
$\omega =\left\langle \omega^{0}\right\rangle$,
and $\rho =\left\langle \rho^{30} \right\rangle$.
From the Lagrangian density (\ref{eq:LRMF}), we derive in the standard way
the equations of motion for the nucleon and meson fields, which are coupled
with each other and can be solved self-consistently.
It is straightforward to obtain the expressions for the free energy density
and pressure in uniform nuclear matter at finite temperature~\citep{Bao16}.

In the Lagrangian density (\ref{eq:LRMF}), an $\omega$-$\rho$ coupling
term (i.e., the last term) is introduced in addition to the original TM1 model.
It is well known that $\omega$-$\rho$ coupling plays a crucial role in
determining the density dependence of the symmetry
energy~\citep{Horo01,Carr03,Cava11,Prov13,Pais16}.
By adjusting the coupling constants, $g_{\rho}$ and $\Lambda_{\rm{v}}$,
it is possible to control the behavior of symmetry energy and its
density dependence. In our previous work~\citep{Bao14b},
we generated a set of RMF models based on the TM1 parametrization,
which have the same isoscalar properties and fixed symmetry energy
at a density of 0.11 fm$^{-3}$ but have different symmetry energy
slope $L$. In the present study, we perform the calculations for pasta phases
employing the extended TM1 model with $L=40$ MeV, which is referred to as the
TM1e model. It is found that the TM1e model provides satisfactory
descriptions for both finite nuclei and neutron stars.
To study the influence of symmetry energy slope $L$, the results of the TM1e model
are compared to those of the original TM1 model with $L=110.8$ MeV.
For completeness, we present in Table~\ref{tab:1} the coupling
constants of the TM1e and TM1 models.
It is shown that only $g_{\rho}$ and $\Lambda_{\rm{v}}$ related
to isovector parts are different, while all other parameters remain the same.
Therefore, the isoscalar saturation properties are the same between these two models,
while the behaviors of symmetry energy are different.
In the TM1e model, the symmetry energy and its slope parameter at saturation density
are $E_{\text{sym}}=31.38$ MeV and $L=40$ MeV,
which are well within the constraints from various observations~\citep{Oert17}.
The corresponding values in the original TM1 model are
$E_{\text{sym}}=36.89$ MeV and $L=110.8$ MeV, which are considered to be rather
large and disfavored by recent astrophysical observations.
\begin{table*}[tbp]
\caption{Coupling constants of the TM1e and TM1 models with symmetry energy $E_{\text{sym}}$
and slope $L$ at saturation density.}
\begin{center}
\begin{tabular}{lcccccccccccccc}
\hline\hline
Model & $E_{\text{sym}}$ (MeV) & $L$ (MeV) & $g_\sigma$  & $g_\omega$ & $g_\rho$ & $g_{2}$ (fm$^{-1}$) & $g_{3}$ & $c_{3}$ & $\Lambda_{\textrm{v}}$ \\
\hline
TM1e  & 31.38 & 40 & 10.0289     & 12.6139    & 13.9714  & $-$7.2325        &0.6183   & 71.3075 & 0.0429  \\
TM1   & 36.89 & 110.8 & 10.0289     & 12.6139    &  9.2644  & $-$7.2325        &0.6183   & 71.3075 & 0.0000  \\
\hline\hline
\end{tabular}
\label{tab:1}
\end{center}
\end{table*}

To describe the pasta phases in hot and dense matter, we employ the CLD
model~\citep{Bao14b,Pais15,Bao16},
where the Wigner--Seitz approximation is adopted for simplifying the
calculation of the free energy.
The nuclear matter inside the Wigner--Seitz cell is assumed to separate
into a dense liquid ($L$) phase and a dilute gas ($G$) phase by a sharp
interface, while the background electron gas is approximated to be uniform
with the density determined by the charge neutrality condition.
In general, the possible geometric structure of pasta phases may change
from droplet to rod, slab, tube, and bubble with increasing baryon density.
At given temperature $T$, average baryon density $n_{b}$, and proton
fraction $Y_{p}$, the equilibrium state can be determined by minimizing
the total free energy density of the system among all configurations
considered~\cite{Latt91,Shen11,Bao14b}.
The free energy density of the pasta phases is expressed as
\begin{eqnarray}
f &=& uf^{L}\left( n_{p}^{L},n_{n}^{L}\right) +\left( 1-u\right) f^{G}\left(
n_{p}^{G},n_{n}^{G},n_{\alpha }^{G}\right)
 \notag \\  & &
+f_{\mathrm{surf}}\left( u,r_{D},\tau \right) +f_{\mathrm{Coul}}\left(
u,r_{D},n_{p}^{L},n_{p}^{G},n_{\alpha }^{G}\right) ,
\label{eq:fws}
\end{eqnarray}
where $u$ is the volume fraction of the liquid phase. The proton and neutron
densities in the liquid (gas) phase are denoted by $n_{p}^{L}$ ($n_{p}^{G}$)
and $n_{n}^{L}$ ($n_{n}^{G}$), respectively.
The free energy contributed from nucleons in phase $i$ ($i=L,G$)
can be calculated in the RMF models~\cite{Shen11,Bao16}.
Note that contributions from electrons are not included in Eq.~(\ref{eq:fws}),
since the background electron gas with a fixed density plays no role in the
minimization procedure. Generally, the contributions from leptons and photons
are treated separately when one constructs the EOS table for astrophysical simulations.
At finite temperature, the $\alpha$ particle may exist as a representative light
nucleus in the dilute gas phase, whereas it is absent in the dense liquid phase.
This is because the $\alpha$ particle tends to dissolve close to nuclear saturation
density due to the finite volume effect~\cite{Latt91,Shen11}.
For simplicity, the $\alpha$ particles are treated as noninteracting Boltzmann
particles in the present calculation.
The surface and Coulomb energy densities are given by
\begin{eqnarray}
{f}_{\mathrm{surf}} &=&\frac{D\tau u_{\mathrm{in}}}{r_{D}},
\label{eq:esurf}
\\
{f}_{\mathrm{Coul}} &=&\frac{e^{2}}{2}\left( \delta n_{c}\right)
^{2}r_{D}^{2}u_{\mathrm{in}}\Phi \left( u_{\mathrm{in}}\right) ,
\label{eq:ecoul}
\end{eqnarray}%
with%
\begin{equation}
\Phi \left( u_{\mathrm{in}}\right) =\left\{
\begin{array}{ll}
\frac{1}{D+2}\left( \frac{2-Du_{\mathrm{in}}^{1-2/D}}{D-2}+u_{\mathrm{in}%
}\right) , & D=1,3 \\
\frac{u_{\mathrm{in}}-1-\ln {u_{\mathrm{in}}}}{D+2}, & D=2.%
\end{array}%
\right.
\end{equation}%
Here, $\tau$ denotes the surface tension, while $D=1,2,3$ is the geometric
dimension of the cell with $r_{D}$ being the size of the inner part.
$u_{\mathrm{in}}$ represents the volume fraction of the inner part,
i.e., $u_{\mathrm{in}}=u$ for droplet, rod, and slab configurations,
and $u_{\mathrm{in}}=1-u$ for tube and bubble configurations.
$e=\sqrt{4\pi /137}$ is the electromagnetic coupling constant.
$\delta n_{c}=n_{p}^{L}-\left(n_{p}^{G}+2n_{\alpha }^{G}\right) $
is the charge-density difference between the liquid and gas phases.
The surface tension $\tau $ is calculated by using the Thomas--Fermi approach
for a one-dimensional nuclear system with the same RMF
parametrization~\cite{Avan10,Bao14a}.
At finite temperature, both the surface energy and
surface entropy are included in the surface tension $\tau $.
It was shown in Ref.~\cite{Bao16} that $\tau $ decreases with increasing
temperature and decreasing proton fraction of the liquid phase.
Meanwhile, it has also been reported that the model with a small slope
parameter $L$ leads to a large surface tension~\cite{Bao14a,Oyam07,Avan12}.

\begin{figure*}[htbp]
 \begin{center}
  \includegraphics[clip,width=12.6 cm]{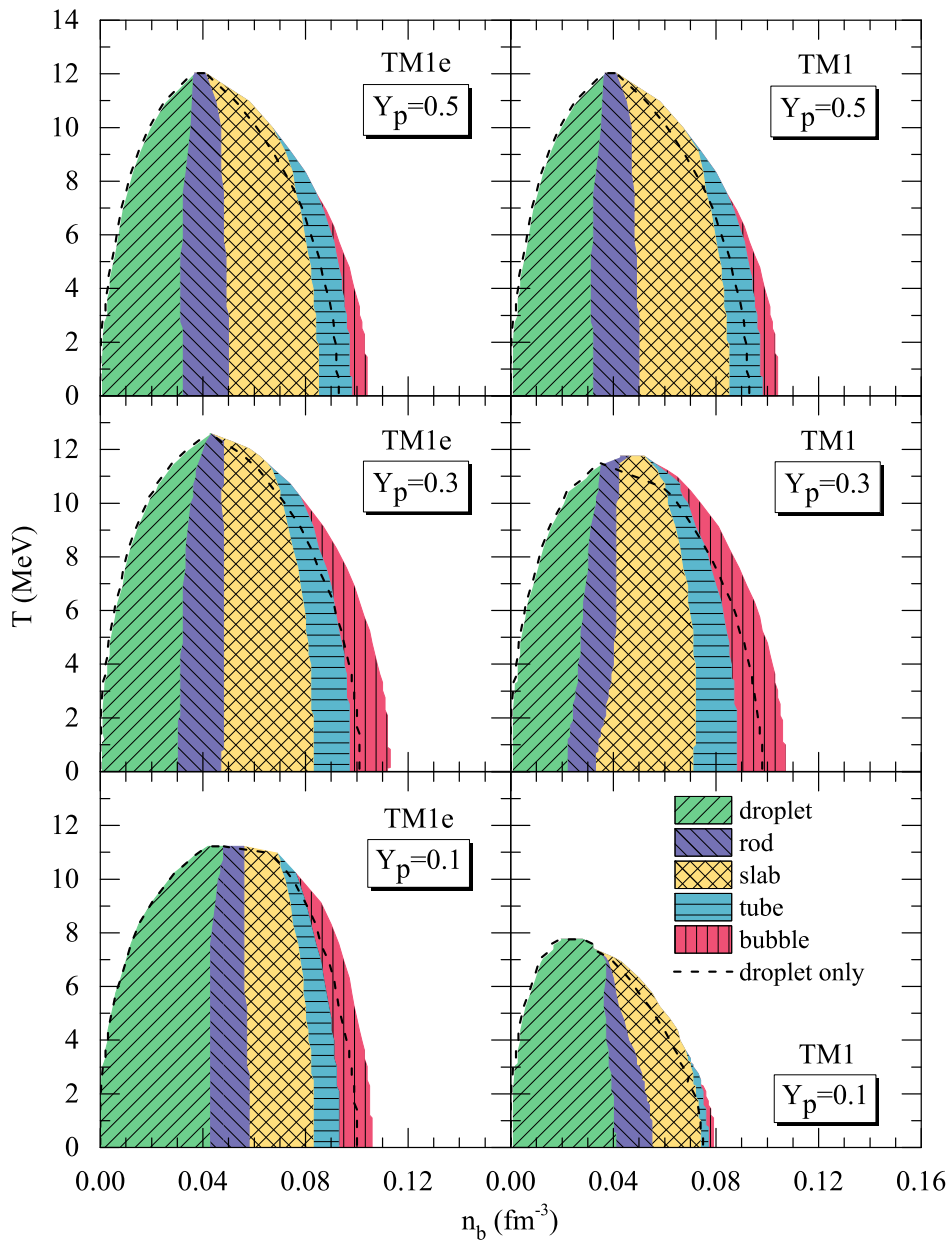}
  \caption{
  Phase diagrams in the $n_b$--$T$ plane for $Y_p=0.1$, $0.3$, and $0.5$
  obtained using the TM1e and TM1 models.
  Different colors indicate the regions for different pasta shapes.
  The boundary of nonuniform matter with only droplet configuration
  is shown by the dashed line for comparison.}
  \label{fig:1Tnb}
 \end{center}
\end{figure*}

\begin{figure*}[htbp]
 \begin{center}
  \includegraphics[clip,width=12.6 cm]{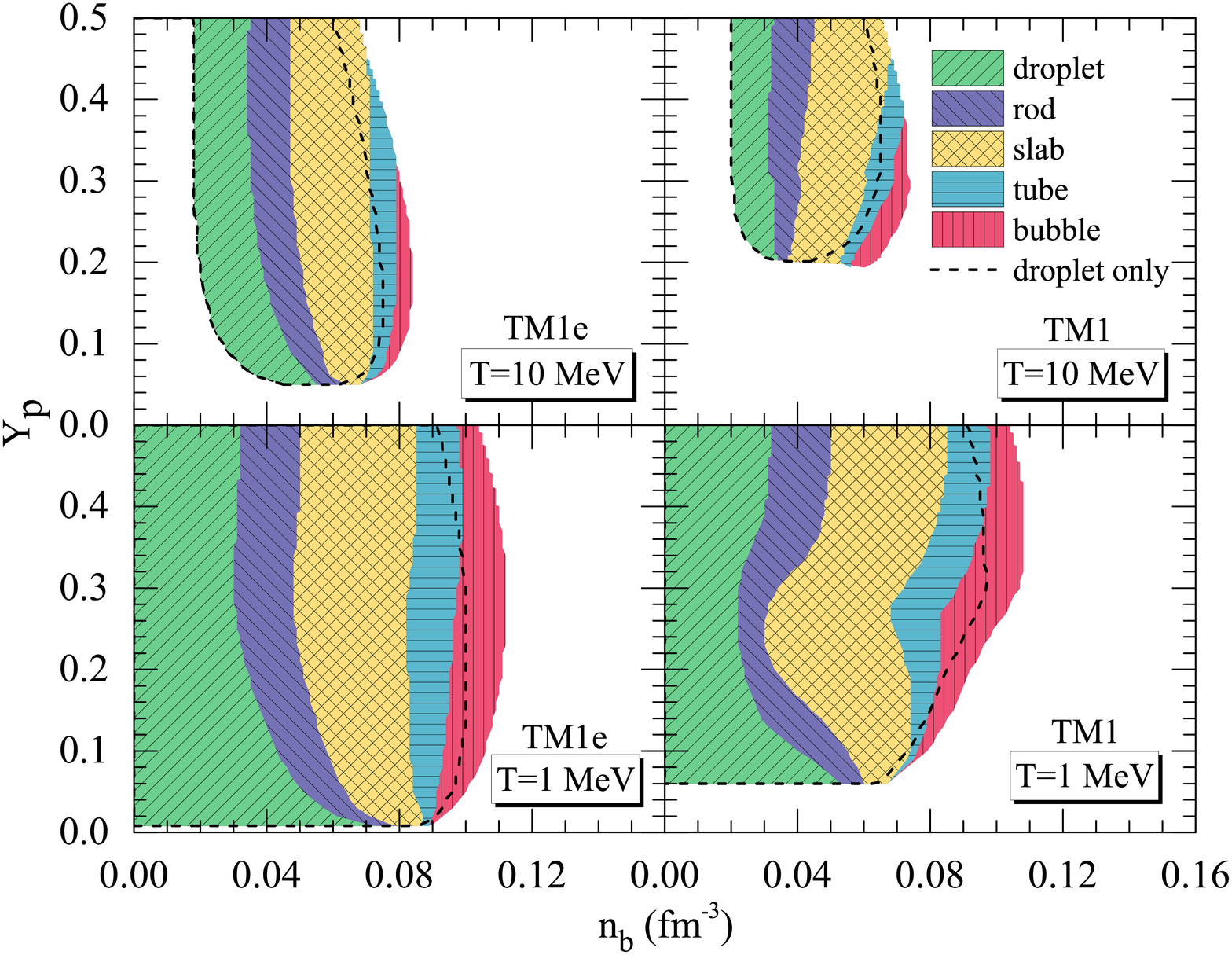}
  \caption{Phase diagrams in the $n_{b}$--$Y_{p}$ plane at $T=1$ and 10 MeV
  obtained using the TM1e and TM1 models.
  Different colors indicate the regions for different pasta shapes.
  The boundary of nonuniform matter with only droplet configuration
  is shown by the dashed line for comparison.}
  \label{fig:2Ypnb}
 \end{center}
\end{figure*}

\begin{figure*}[htbp]
 \begin{center}
  \includegraphics[clip,width=12.6 cm]{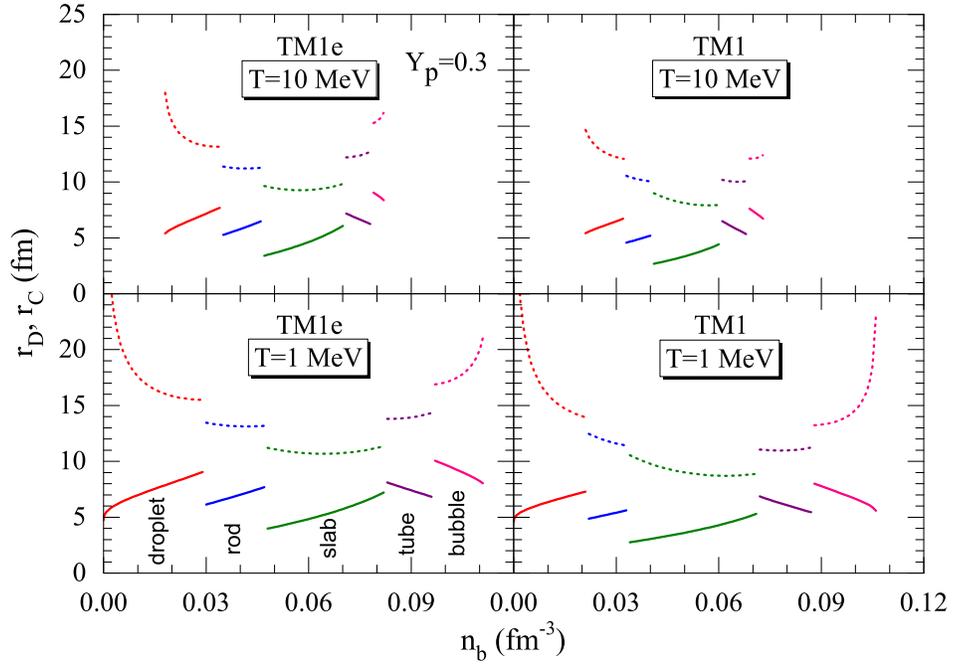}
  \caption{Size of the nuclear pasta, $r_D$ (solid lines), and that of the Wigner--Seitz
  cell, $r_C$ (dotted lines), as a function of the baryon density $n_b$
  obtained using the TM1e and TM1 models.
  The results for $Y_p=0.3$ at $T=1$ and $10$ MeV are shown in the lower and upper
  panels, respectively. }
  \label{fig:3rnb}
 \end{center}
\end{figure*}

\begin{figure}[htbp]
 \begin{center}
  \includegraphics[clip,width=8.6 cm]{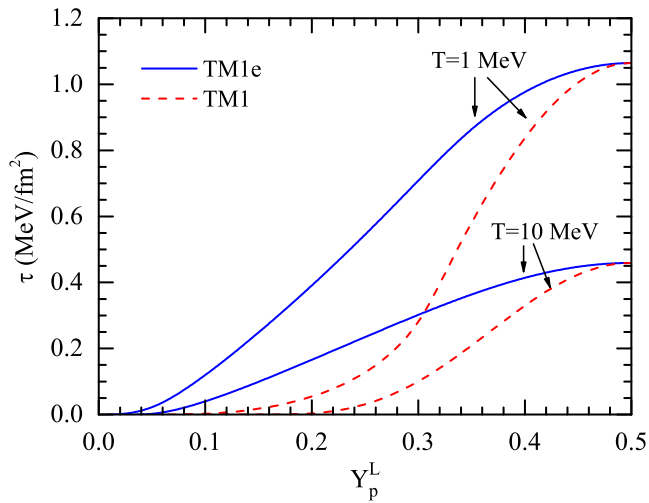}
  \caption{Surface tension, $\tau$, as a function of the proton fraction in the
  liquid phase, $Y^L_{p}$, at $T=1$ and $10$ MeV.
  The results are calculated from the Thomas--Fermi approach for a one-dimensional
  nuclear system using the TM1e and TM1 models. }
  \label{fig:4tao}
 \end{center}
\end{figure}

\begin{figure*}[htbp]
\begin{center}
\begin{tabular}{cc}
  \includegraphics[clip,width=8.6cm]{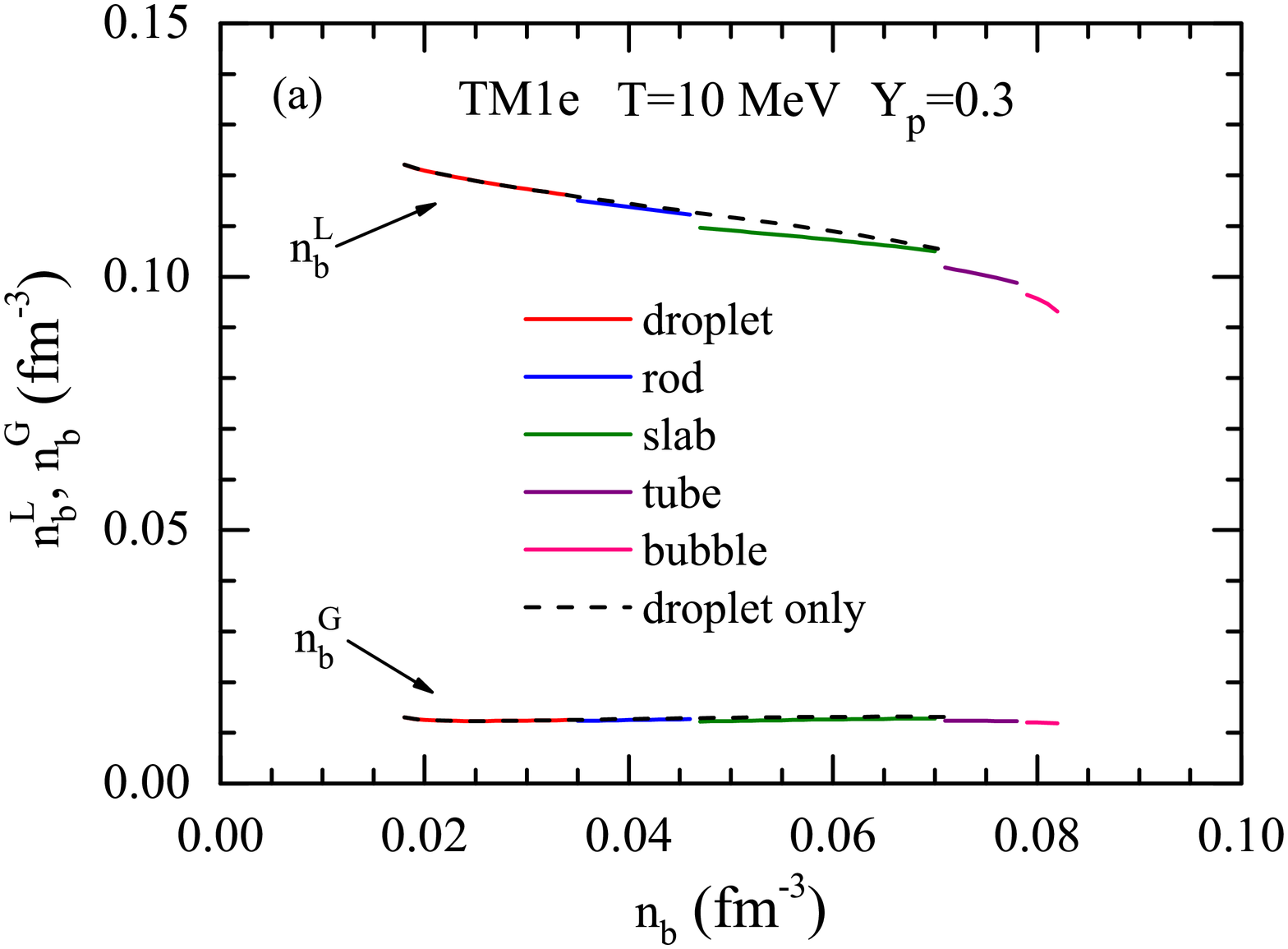} &
  \includegraphics[clip,width=8.6cm]{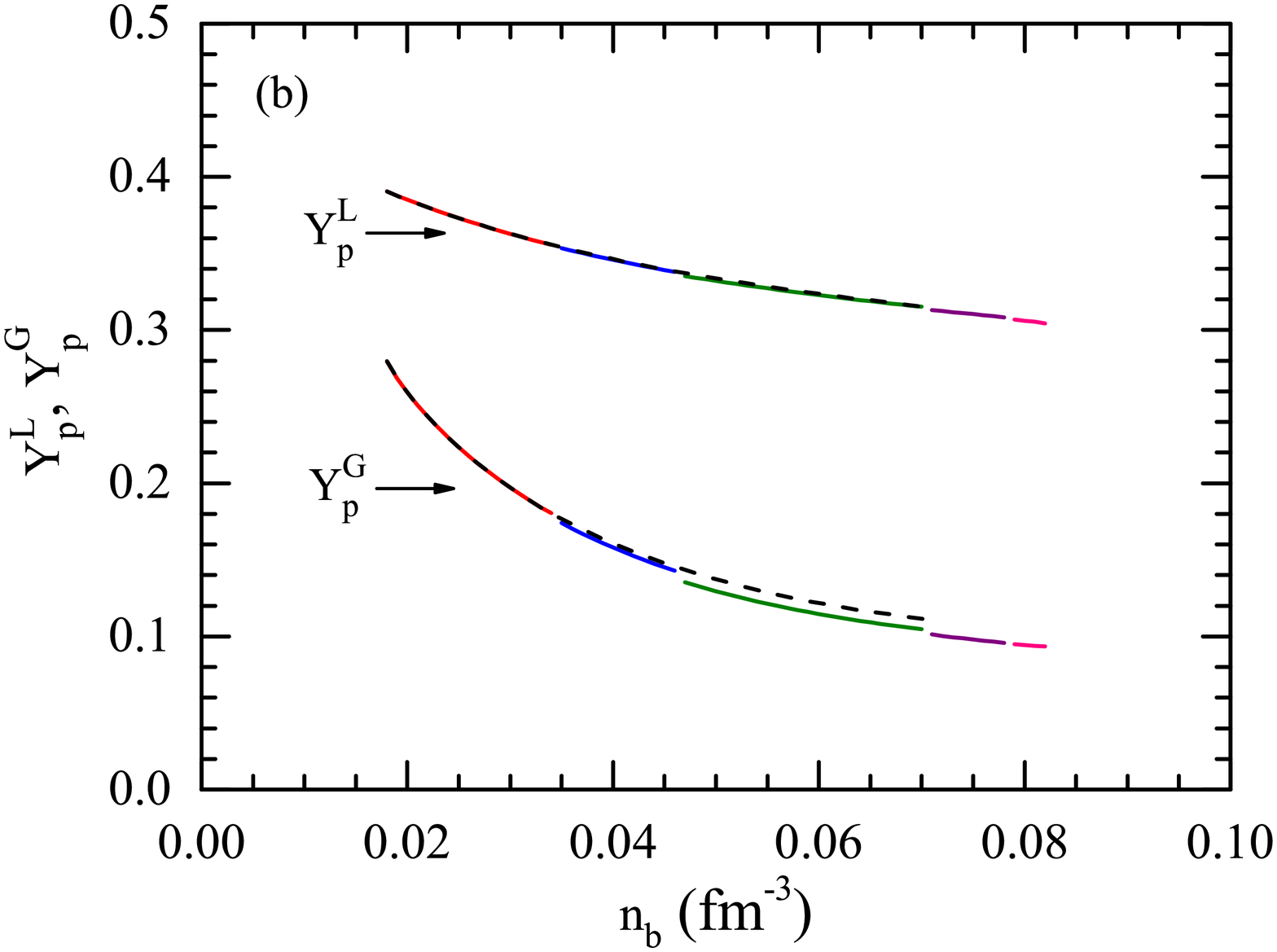} \\
  \includegraphics[clip,width=8.6cm]{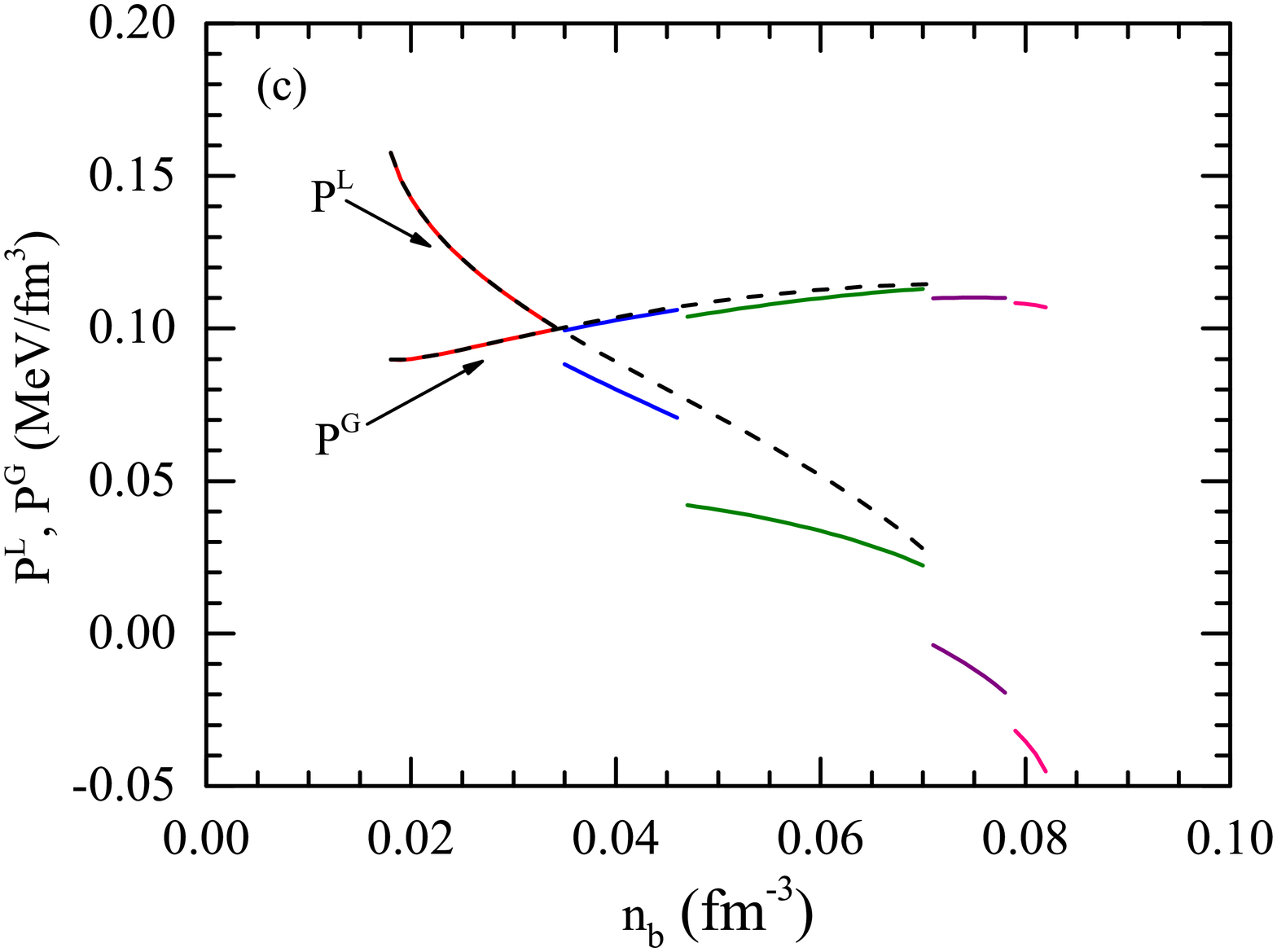} &
  \includegraphics[clip,width=8.6cm]{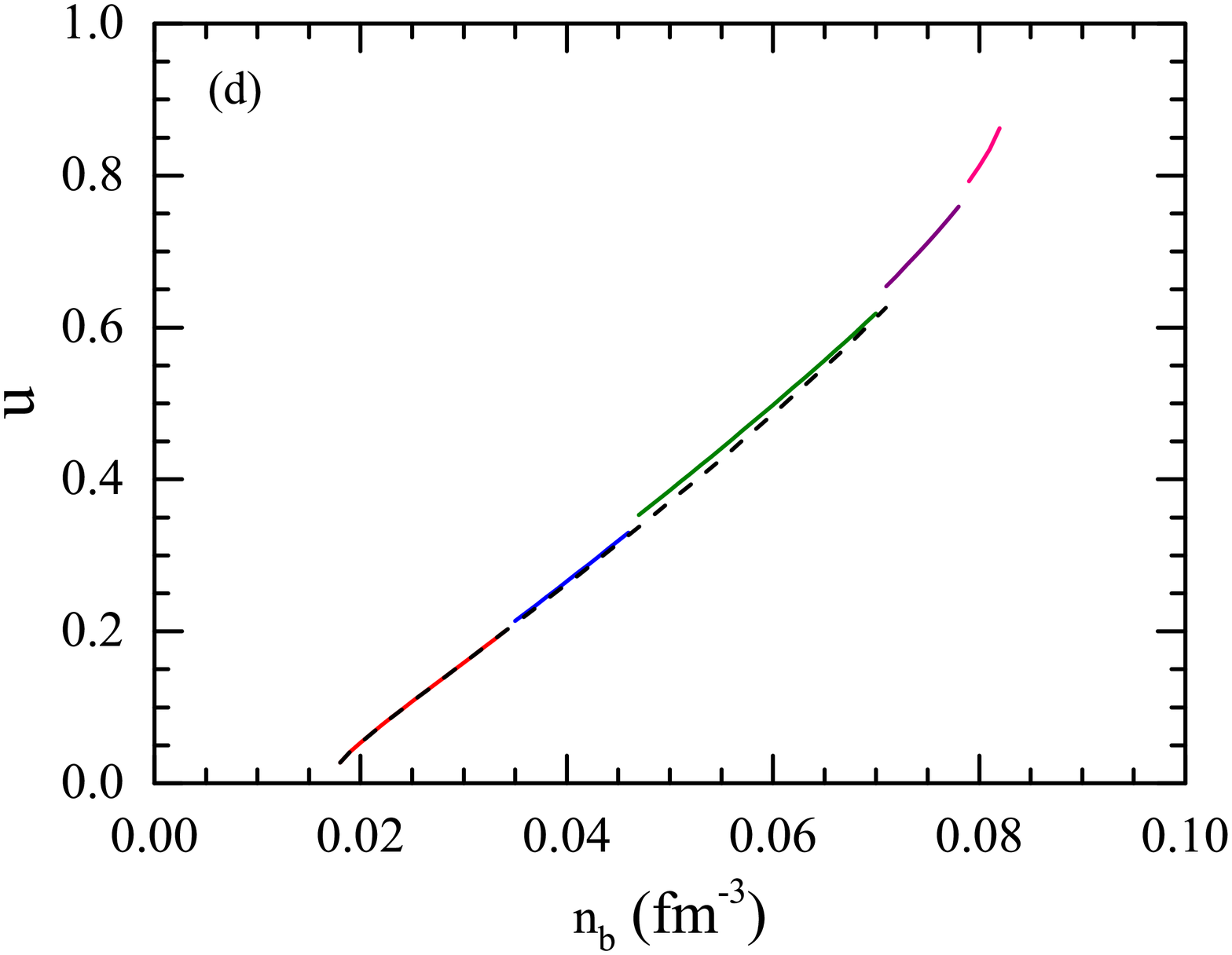} \\
\end{tabular}
  \caption{Properties of the pasta phases at $T = 10$ MeV and $Y_{p} = 0.3$ using the TM1e model. The (a) coexisting liquid and gas densities $n^L_{b}$ and $n^G_{b}$, (b) proton fractions $Y^L_{p}$ and $Y^G_{p}$, (c) pressures $P^L$ and $P^G$, and (d) volume fraction of the liquid phase, $u$, are plotted as a function of the average baryon density $n_{b}$.}
  \label{fig:5CLD}
 \end{center}
\end{figure*}

\begin{figure}[htbp]
 \begin{center}
  \includegraphics[clip,width=8.6 cm]{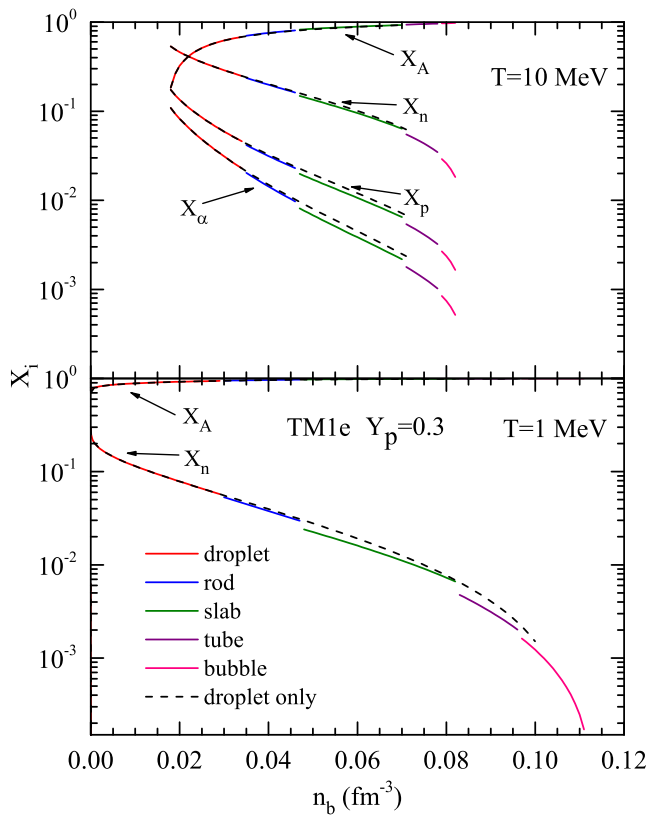}
  \caption{Fractions of neutrons ($X_n$), protons ($X_p$), $\alpha$ particles ($X_\alpha$), and
  heavy nuclei ($X_A$) as a function of the average baryon density $n_{b}$ in nonuniform matter
  for $Y_p=0.3$ at $T=1$ and $10$ MeV using the TM1e model. The results with nuclear
  pasta (solid lines) are compared to those with droplet only (dashed lines). }
  \label{fig:6Xi}
 \end{center}
\end{figure}

\begin{figure}[htbp]
 \begin{center}
  \includegraphics[clip,width=8.6 cm]{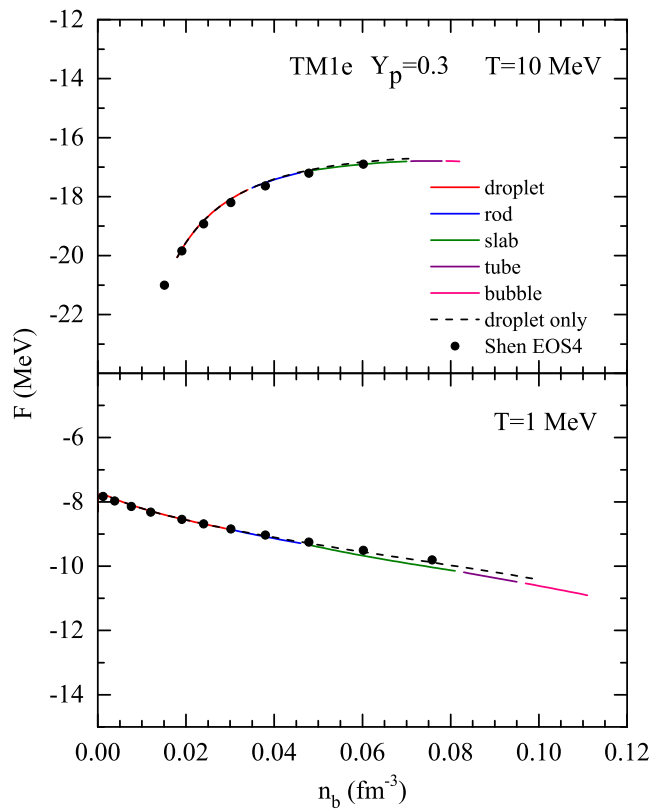}
  \caption{Free energy per baryon $F$ as a function of the baryon density $n_{b}$
  for $Y_{p}=0.3$ at $T=1$ and $10$ MeV using the TM1e model.
  The results with nuclear pasta (solid lines) are compared to those with droplet only (dashed lines).
  The dots represent the values from the Shen EOS4~\citep{Shen20}, which are
  calculated by a parameterized Thomas--Fermi approximation.}
  \label{fig:7F}
 \end{center}
\end{figure}

\begin{figure}[htbp]
 \begin{center}
  \includegraphics[clip,width=8.6 cm]{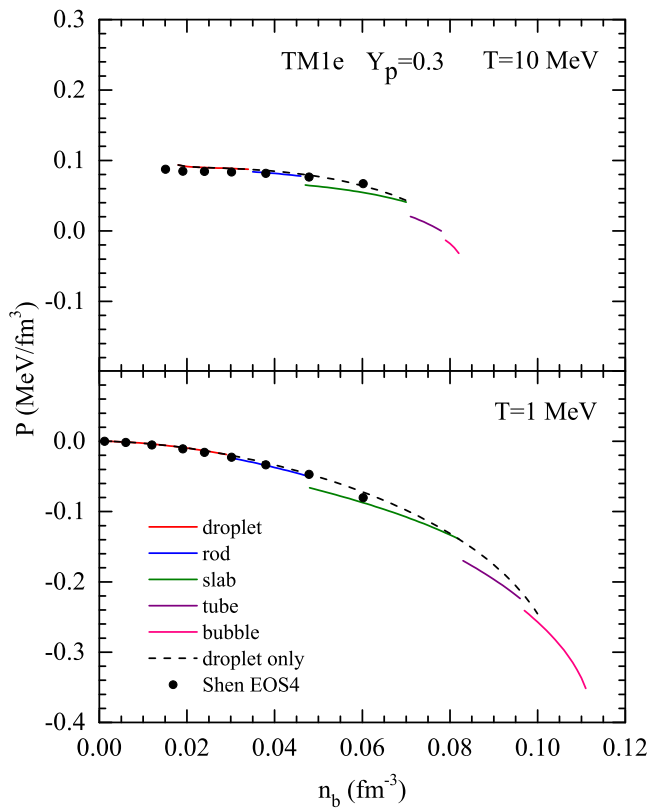}
  \caption{Same as Fig.~\ref{fig:7F}, but for pressure $P$.}
  \label{fig:8P}
 \end{center}
\end{figure}

\begin{figure}[htbp]
 \begin{center}
  \includegraphics[clip,width=8.6 cm]{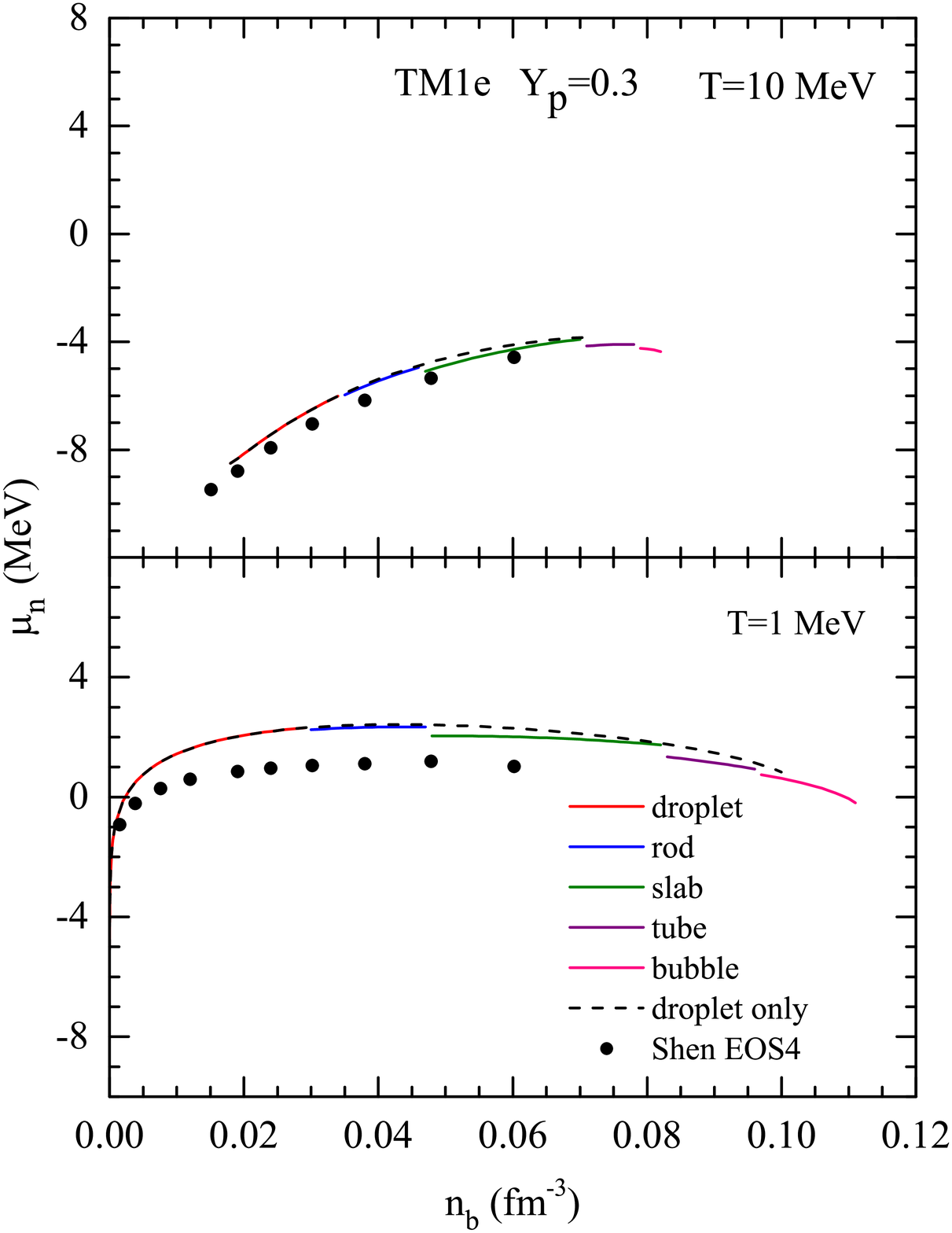}
  \caption{Same as Fig.~\ref{fig:7F}, but for chemical potential of neutrons, $\mu_{n}$.}
  \label{fig:9mun}
 \end{center}
\end{figure}

\begin{figure}[htbp]
 \begin{center}
  \includegraphics[clip,width=8.6 cm]{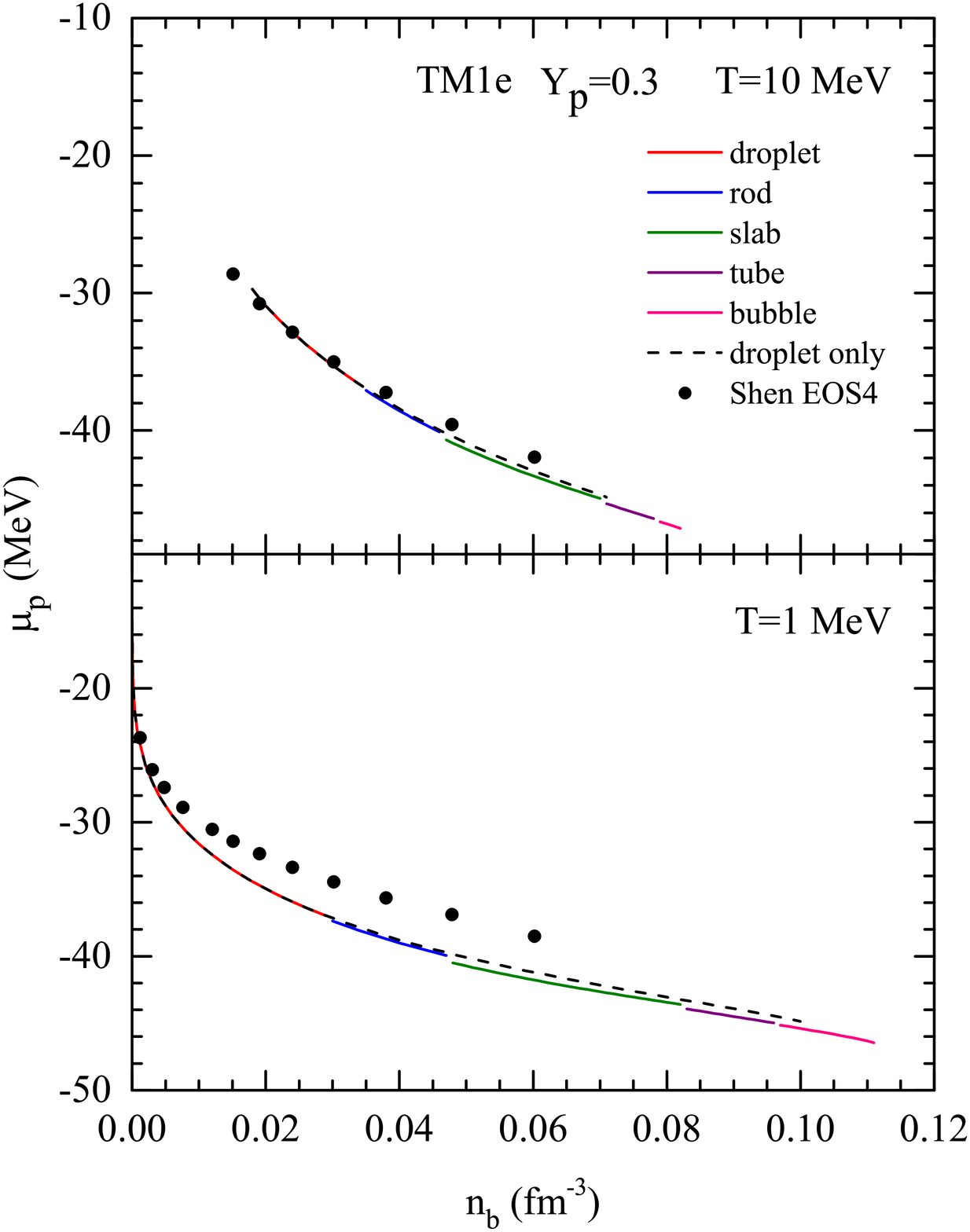}
  \caption{Same as Fig.~\ref{fig:7F}, but for chemical potential of protons, $\mu_{p}$.}
  \label{fig:10mup}
 \end{center}
\end{figure}

With given average density $n_{b}$ and proton fraction $Y_{p}$,
the free energy density $f$ in Eq.~(\ref{eq:fws}) is
considered a function of seven variables:
$n_{p}^{L}$, $n_{n}^{L}$, $n_{p}^{G}$, $n_{n}^{G}$, $n_{\alpha }^{G}$, $u$, and $r_{D}$.
These variables satisfy the constraints of the proton and neutron number conservation,
which can be expressed as
\begin{eqnarray}
un_{p}^{L}+\left( 1-u\right) \left( n_{p}^{G}+2n_{\alpha }^{G}\right)
&=&n_{b}Y_{p},
\label{eq:cnp} \\
un_{n}^{L}+\left( 1-u\right) \left( n_{n}^{G}+2n_{\alpha }^{G}\right)
&=&n_{b}\left( 1-Y_{p}\right).
\label{eq:cnn}
\end{eqnarray}%
In order to derive the phase equilibrium conditions by minimizing the free energy
density, we introduce the Lagrange multipliers $\mu _{p}$ and $\mu_{n}$ for the constraints,
and then perform the minimization for the function,
\begin{eqnarray}
w &=& f-\mu _{p}\left[ un_{p}^{L}+\left( 1-u\right) \left( n_{p}^{G}+2n_{\alpha
}^{G}\right) \right]
 \notag \\  & &
-\mu _{n}\left[ un_{n}^{L}+\left( 1-u\right) \left(
n_{n}^{G}+2n_{\alpha }^{G}\right) \right].
\end{eqnarray}
By minimizing $w$ with respect to the variables, we obtain the following relations:
\begin{eqnarray}
0 &=&\frac{\partial w}{\partial n_{n}^{L}}
   =u\left[ \mu _{n}^{L}-\mu _{n}\right] ,
\label{eq:min1} \\
0 &=&\frac{\partial w}{\partial n_{n}^{G}}
   =(1-u)\left[ \mu _{n}^{G}-\mu _{n}\right],
\label{eq:min2} \\
0 &=&\frac{\partial w}{\partial n_{p}^{L}}
   =u\left[ \mu _{p}^{L}-\mu _{p}\right] +\frac{2f_{\mathrm{Coul}}}{\delta n_{c}},
\label{eq:min3} \\
0 &=&\frac{\partial w}{\partial n_{p}^{G}}
   =(1-u)\left[ \mu _{p}^{G}-\mu _{p}\right] -\frac{2{f}_{\mathrm{Coul}}}{\delta n_{c}},
\label{eq:min4} \\
0 &=&\frac{\partial w}{\partial n_{\alpha }^{G}}
  =(1-u)\left[ \mu _{\alpha}^{G}-2\left( \mu _{p}+\mu _{n}\right) \right]
     -\frac{4{f}_{\mathrm{Coul}}}{\delta n_{c}},
\label{eq:min5} \\
0 &=&\frac{\partial w}{\partial u}
  =\left[ f^{L}-\mu _{p}n_{p}^{L}-\mu_{n}n_{n}^{L}\right] \notag\\
  & & \hspace{0.6cm}
-\left[ f^{G}-\mu _{p}\left( n_{p}^{G}+2n_{\alpha
}^{G}\right) -\mu _{n}\left( n_{n}^{G}+2n_{\alpha }^{G}\right) \right] \notag\\
  & & \hspace{0.6cm}\pm \left[ \frac{f_{\mathrm{surf}}}{u_{\mathrm{in}}}+\frac{f_{\mathrm{Coul}}}{u_{%
\mathrm{in}}}\left( 1+u_{\mathrm{in}}\frac{\Phi ^{^{\prime }}}{\Phi }\right) %
\right] ,
\label{eq:min6} \\
0 &=&\frac{\partial w}{\partial r_{D}}
  =-\frac{f_{\mathrm{surf}}}{r_{D}}+%
\frac{2f_{\mathrm{Coul}}}{r_{D}}.
\label{eq:min7}
\end{eqnarray}
According to Eqs.~(\ref{eq:min1})--(\ref{eq:min5}), the equilibrium conditions for chemical
potentials are written as
\begin{eqnarray}
\mu _{n}^{G} &=&\mu _{n}^{L},
\label{eq:cmun} \\
\mu _{p}^{G} &=&\mu _{p}^{L}+\frac{2f_{\mathrm{Coul}}}{u(1-u)\delta n_{c}},
\label{eq:cmup} \\
\mu _{\alpha }^{G} &=&2\mu _{p}^{G}+2\mu _{n}^{G} .
\label{eq:cmua}
\end{eqnarray}%
The equilibrium condition for the pressures between the liquid and gas phases is
achieved from Eq.~(\ref{eq:min6}) and written as
\begin{eqnarray}
P^{G} &=&P^{L}+\frac{2f_{\mathrm{Coul}}}{\delta n_{c}}\left( \frac{n_{p}^{L}%
}{u}+\frac{n_{p}^{G}+2n_{\alpha }^{G}}{1-u}\right) \notag\\
  & & \mp \frac{f_{\mathrm{Coul}}}{u_{\mathrm{in}}}
  \left( 3+u_{\mathrm{in}}\frac{\Phi ^{^{\prime }}}{\Phi }%
\right) ,
\label{eq:cp}
\end{eqnarray}%
where the sign of the last term is \textquotedblleft $-$\textquotedblright\ for droplet,
rod, and slab configurations, or \textquotedblleft $+$\textquotedblright\ for tube
and bubble configurations.
It is clear that equilibrium conditions for two-phase coexistence are
altered due to the inclusion of surface and Coulomb terms in the minimization
procedure and, as a result, they are different from the Gibbs equilibrium conditions.
Compared to the Gibbs conditions with equal pressures and chemical potentials
between the two phases, the additional terms in Eqs.~(\ref{eq:cmup}) and (\ref{eq:cp})
are caused by the surface and Coulomb contributions.
If we neglect the finite-size effects by taking the limit $\tau \rightarrow 0$,
these additional terms disappear and the equilibrium equations would reduce to the
Gibbs conditions.
Based on the equilibrium condition $f_{\mathrm{surf}}=2f_{\mathrm{Coul}}$
from Eq.~(\ref{eq:min7}), the size of the inner phase and that of the Wigner--Seitz
cell are respectively given by
\begin{eqnarray}
r_{D} &=&\left[ \frac{\tau {D}}{e^{2}\left( \delta n_{c}\right) ^{2}\Phi }%
\right] ^{1/3},
\label{eq:rD} \\
r_{C} &=&u_{\mathrm{in}}^{-1/D}r_{D}.
\label{eq:rC}
\end{eqnarray}

At given temperature $T$, average baryon density $n_{b}$, and proton
fraction $Y_{p}$, we solve the equilibrium conditions together with
the coupled equations of the RMF model in the liquid and gas phases
for all pasta shapes, and then determine the thermodynamically stable
state that has the lowest free energy density.
In the pasta phases, the pressure and chemical potentials of the system
may be different from those in the liquid and gas phases.
Therefore, we compute these quantities by the thermodynamic relations
\begin{eqnarray}
 P &=& \left[ n_b^{2} \frac{\partial f \left(T,Y_{p},n_{b}\right)/n_{b}}{\partial n_b} \right]_{T,Y_{p}},
\label{eq:ppre} \\
\mu_{p} &=& \left[\frac{\partial f \left(T,Y_{p},n_{b}\right)}{\partial n_p} \right]_{T,n_n},
\label{eq:pmup} \\
\mu_{n} &=& \left[\frac{\partial f \left(T,Y_{p},n_{b}\right)}{\partial n_n} \right]_{T,n_p},
\label{eq:pmun}
\end{eqnarray}
where $n_{p}=Y_{p}n_{b}$ and $n_{n}=\left(1-Y_{p}\right)n_{b}$ are the average number densities of protons and neutrons, respectively.

\section{Results and discussion}
\label{sec:3}

We explore the properties of nuclear pasta and its influence on
the EOS for astrophysical simulations. The pasta phases are calculated
in the CLD model, where a nuclear liquid coexists with a dilute gas
consisting of free nucleons and $\alpha$ particles employing a sharp interface.
For the nuclear interaction, we employ the TM1e model with a small
symmetry energy slope $L=40$ MeV, which is compatible with both
experimental nuclear data and recent observations of neutron stars.
To evaluate the effects of nuclear symmetry energy on the pasta phases,
we compare the results of TM1e to those of the original TM1 model with
a large symmetry energy slope $L=110.8$ MeV. The difference between the
TM1e and TM1 models is only in the isovector part, while the isoscalar
properties in the two models remain the same.

We first discuss the phase diagram of hot and dense matter including
nuclear pasta. At given temperature $T$, proton fraction $Y_p$, and
baryon number density $n_b$, we perform calculations for all pasta phases,
and then determine the most stable shape among them with the lowest free energy
density. The transition to uniform matter occurs at a density of $n_t$ where
the free energy density of homogeneous matter becomes lower than that of pasta phases.
In Fig.~\ref{fig:1Tnb}, we show the phase diagrams in the $n_b$--$T$
plane for $Y_p=0.1$, $0.3$, and $0.5$ obtained in the TM1e model (left panels)
compared to that in the TM1 model (right panels).
The results with only droplet configuration are plotted by the dashed lines,
so that the influence of nuclear pasta on the phase diagram can be estimated.
It is found that the inclusion of pasta phases delays the transition to
uniform matter. This is because the configuration space is enlarged by
considering nonspherical nuclei in addition to the droplet.
One can see that the density range of nonuniform matter depends on both $T$ and $Y_p$.
At low temperatures, various pasta shapes appear one by one with increasing density,
and the transition between different shapes is only weakly dependent on $T$.
As the temperature increases, the density range of nonuniform matter shrinks,
while some pasta shapes like bubble and tube may not occur before the transition to
uniform matter. Eventually, the temperature reaches the critical value $T_c$ where
the nonuniform matter phase disappears completely; i.e., nuclear pasta cannot be
formed at $T>T_c$.
Clearly, the critical temperature $T_c$ for $Y_p=0.1$ obtained in the TM1 model
is much smaller than in other cases. This is because the TM1 model has a rather large
symmetry energy slope $L=110.8$ MeV and a large $L$ is generally correlated to
a small crust-core transition density in neutron stars~\citep{Bao15}.
By comparing the results of the TM1e model (left panels) to those of the TM1
model (right panels), one can see the influence of the symmetry energy slope
on the phase diagram.
There is almost no difference in the case of $Y_p=0.5$ and the difference
for $Y_p=0.3$ is still small. However, a significant difference between
the TM1e and TM1 models is observed in the case of $Y_p=0.1$.
This is because the two models have the same isoscalar properties but different symmetry
energy behavior. It is well known that the symmetry energy plays an important role
in neutron-rich matter, but it has no impact on the properties of symmetric nuclear matter.
A similar effect of the symmetry energy slope on the phase diagram was also reported
in Refs.~\cite{Toga17,Shen20}, where the parametrized Thomas--Fermi approximation
was used and only spherical nuclei were taken into account.
In Fig.~\ref{fig:2Ypnb}, we show the phase diagrams in the $n_b$--$Y_p$
plane for $T=1$ and $10$ MeV obtained in the TM1e and TM1 models.
It is seen that the onset of various pasta shapes is somewhat dependent on $Y_p$.
There are significant differences between the TM1e and TM1 models in the low-$Y_p$ region,
where the behavior of symmetry energy plays a crucial role.
It is found that nuclear pasta cannot be formed in the TM1 model for $Y_p<0.2$ at $T=10$ MeV,
whereas it exists until $Y_p\approx0.05$ in the TM1e model.
At $T=1$ MeV, the region of nuclear pasta extends to a lower value of $Y_p$ compared to the
case of $T=10$ MeV.

It is interesting to investigate the properties of nuclear pasta appearing in nonuniform matter.
We show in Fig.~\ref{fig:3rnb} the size of the nuclear pasta ($r_D$) and that of the Wigner--Seitz
cell ($r_C$) as a function of the baryon density $n_b$.
The results of the TM1e model (left panels) for $Y_p=0.3$ at $T=1$ and $10$ MeV are compared to
those obtained in the TM1 model (right panels). It is observed that $r_D$ in the droplet, rod,
and slab phases increases with increasing $n_b$, whereas $r_D$ in the tube and bubble phases
decreases. This is related to an increase of the volume fraction of the liquid phase.
There are obvious discontinuities in $r_D$ and $r_C$ at the transition between
different pasta shapes, which exhibit the character of the first-order transition.
Comparing the results between the TM1e and TM1 models, the tendencies of $r_D$ and $r_C$
are very similar in the two models. It is noticed that the value of $r_D$ in the TM1e model
is slightly larger than that in the TM1 model. This is mainly due to the difference of the
surface tension $\tau$, which is displayed in Fig.~\ref{fig:4tao}.
According to Eq.~(\ref{eq:rD}), a large surface tension $\tau$ generally leads to a large
nuclear size $r_D$. From Fig.~\ref{fig:4tao}, we can see that the TM1e model with a small
symmetry energy slope $L=40$ MeV predicts much larger surface tension than the TM1 model
with $L=110.8$ MeV. The correlation between the slope $L$ and the surface tension $\tau$
has also been discussed in Refs.~\citep{Oyam07,Avan12,Bao16}.

In Fig.~\ref{fig:5CLD}, we present several properties of nuclear pasta described in
the CLD model, where the liquid phase with density $n_{b}^{L}$ and proton fraction
$Y_{p}^{L}$ coexists with
the gas phase with $n_{b}^{G}$ and $Y_{p}^{G}$.
The equilibrium conditions for two-phase coexistence are given by
Eqs.~(\ref{eq:cmun})--(\ref{eq:cp}).
We plot in Fig.~\ref{fig:5CLD} the following quantities as a function of
the average baryon density $n_b$:
the coexisting liquid and gas densities $n^L_{b}$ and $n^G_{b}$ [Fig.~\ref{fig:5CLD}(a)],
proton fractions $Y^L_{p}$ and $Y^G_{p}$ [Fig.~\ref{fig:5CLD}(b)],
pressures $P^L$ and $P^G$ [Fig.~\ref{fig:5CLD}(c)],
and volume fraction of the liquid phase $u$ [Fig.~\ref{fig:5CLD}(d)].
The calculations are performed at $T=10$ MeV and $Y_p=0.3$ with the TM1e model.
To explore the differences between spherical and nonspherical nuclei, we show
the results with only droplet configuration by the dashed lines.
It is found that the differences between pasta phases and droplet configuration
are rather small in Figs.~\ref{fig:5CLD}(a), ~\ref{fig:5CLD}(b), and ~\ref{fig:5CLD}(d), whereas considerable differences
are observed in the pressure of Fig.~\ref{fig:5CLD}(c).
Since the surface and Coulomb contributions are distinguished for different pasta shapes,
it leads to the jumps at the transition between pasta shapes.
It is noticeable that the pressure of the liquid phase ($P^L$) is clearly different
from that of the gas phase ($P^G$), which is due to the surface and Coulomb
contributions given by the last two terms in Eq.~(\ref{eq:cp}).
On the contrary, the coexisting liquid and gas phases have equal
pressures according to the Gibbs equilibrium conditions used in the CP method.
In Fig.~\ref{fig:5CLD}(b), the proton fraction $Y^L_{p}$ decreases with increasing
$n_b$, which implies that heavy nuclei become more neutron rich before dissolving
into uniform matter. The volume fraction of the liquid phase $u$ shown in
Fig.~\ref{fig:5CLD}(d) increases monotonically and nonspherical nuclei appear at
$u\approx0.21$. A simple estimate based on the Bohr-Wheeler fission condition indicates
that a spherical nucleus becomes unstable to quadrupolar deformation
at $u > 1/8$~\citep{Peth95}.
In the CP method, the transition from droplet to rod occurs at $u\approx0.22$,
which is very close to the value obtained in the present calculation using the CLD method.

We display in Fig.~\ref{fig:6Xi} the fractions of neutrons, protons, $\alpha$ particles,
and heavy nuclei as a function of the average baryon density $n_b$ in nonuniform matter
for $Y_p=0.3$ at $T=1$ and $10$ MeV. These quantities are calculated in the CLD model
by $X_A=u \left( n_{n}^{L}+n_{p}^{L}\right)/n_b$, $X_\alpha=(1-u) 4n_{\alpha}^{G}/n_b$,
and $X_i=(1-u) n_{i}^{G}/n_b\; (i=n,p)$.
Compared to the results with only droplet configuration (dashed lines),
the tendency of $X_i$ with the inclusion of nuclear pasta is very similar,
but small discontinuities appear at the change of pasta shapes.
In the case of $T=10$ MeV (upper panel), there are noticeable fractions of protons ($X_p$)
and $\alpha$ particles ($X_\alpha$), which are reduced to almost zero at
$T=1$ MeV (lower panel). Moreover, the fraction of heavy nuclei ($X_A$) is dominant
in nonuniform matter and the value of $X_A$ at $T=10$ MeV is smaller than that at $T=1$ MeV.
This is because, at higher temperature, the particle densities in the gas phase are
significantly enhanced, whereas the densities in the liquid phase are insensitive to the
temperature due to their high degeneracy.
We can see that $X_n$, $X_p$, and $X_\alpha$ decrease with increasing $n_b$,
which is related to the increase of the volume fraction $u$
shown in Fig.~\ref{fig:5CLD}(d).

It is essential to analyze the influence of nuclear pasta on the thermodynamic quantities
which play crucial roles in numerical simulations of core-collapse supernovas and
neutron-star mergers. It is also important to compare the present results in the CLD model
to those from a realistic EOS table with the same nuclear interaction, so that the
uncertainty due to different descriptions of nonuniform matter can be estimated.
In Fig.~\ref{fig:7F}, we show the free energy per baryon, $F$, as a function of the
baryon density $n_b$ for $Y_p=0.3$ at $T=1$ and $10$ MeV.
The results with the inclusion of nuclear pasta (solid lines) are slightly smaller than
those with droplet only (dashed lines) due to the enlargement of the configuration space
by considering nonspherical nuclei. Meanwhile, the results from the realistic Shen EOS4~\citep{Shen20},
which were constructed using a parametrized Thomas--Fermi approximation with the TM1e model,
are shown by dots for comparison. It is found that the values of $F$ taken from
the Shen EOS4 are very close to the present results obtained using the CLD method.
This confirms that the two methods are consistent with each other for calculating the free energies.
We see that $F$ increases with the density in the case of $T=10$ MeV (upper panel),
while it decreases at $T=1$ MeV (lower panel). This is because $F=E-TS$ is related to
the behaviors of the internal energy $E$ and the entropy $S$.
As the density increases, the entropy per baryon, $S$, decreases
(see, e.g., Fig. 11 of Ref.~\cite{Shen20}), which leads to the increase of $F$
at higher temperature. On the contrary, the entropy plays less of a role at lower temperature,
where the decrease of internal energy $E$ is dominant.
In Fig.~\ref{fig:8P}, we display the pressure $P$ as a function of the
baryon density $n_b$ for $Y_p=0.3$ at $T=1$ and $10$ MeV.
Compared to the results with droplet only (dashed lines),
small discontinuities are observed in the pressures with the inclusion
of pasta phases (solid lines) due to the change of pasta shapes.
The discontinuities exhibit the character of the first-order transition.
It is found that the pressures taken from the Shen EOS4 are consistent with
the present results obtained using the CLD method.
Comparing the cases between $T=1$ and $10$ MeV, the pressure at higher temperature
is relatively larger, while the tendencies of $P$ in the two cases are very similar.
We note that the result shown in Fig.~\ref{fig:8P} represents the baryon pressure without
contributions from electrons and photons. In fact, the pressure of nonuniform matter
is dominated by the background electron gas, which ensures the total pressure is positive.
Therefore, the influence of nuclear pasta on the pressure is neglectable.
In Figs.~\ref{fig:9mun} and~\ref{fig:10mup}, the chemical potentials of neutrons and protons,
$\mu_{n}$ and $\mu_{p}$, are shown as a function of the baryon density $n_b$
for $Y_p=0.3$ at $T=1$ and $10$ MeV.
It is observed that the results with nuclear pasta (solid lines) are very close to those
with droplet only (dashed lines), whereas the change of pasta shapes may cause small
discontinuities in $\mu_{n}$ and $\mu_{p}$.
One can see that $\mu_{p}$ decreases with increasing $n_b$, which is related to the
decrease of the proton density in the gas phase.
There are visible differences between the present results of the CLD model and those from
the Shen EOS4 obtained by a Thomas--Fermi calculation.
This implies that the chemical potentials are relatively sensitive to the method used for
describing nonuniform matter.
Since the chemical potentials are calculated from the first derivative of the free energy
as given in Eqs.~(\ref{eq:pmup}) and (\ref{eq:pmun}), the differences in chemical potentials
could be more obvious than those in the free energy.
Furthermore, the chemical potentials, $\mu_{n}$ and $\mu_{p}$, are sensitively dependent
on the density distributions of protons and neutrons, which are clearly different between
the CLD model and the parametrized Thomas--Fermi approximation.
The relatively large differences in $\mu_{p}$ between the present results and those from
the Shen EOS4 may be partly due to different treatments of the Coulomb contributions between
the two methods. In the present calculation using the CLD method, the Coulomb energy
is related to the surface energy by the equilibrium condition $f_{\mathrm{surf}}=2f_{\mathrm{Coul}}$,
where the surface tension $\tau$ is determined self-consistently as described
in Refs.~\cite{Avan10,Bao14a}.
In the parametrized Thomas--Fermi approximation used in the Shen EOS4,
the Coulomb energy is related to a gradient parameter $F_0$, which is somewhat
underestimated in comparison to the self-consistent Thomas--Fermi approximation~\cite{Zhang14}.
Notable differences in $\mu_{p}$ were also found and discussed in Ref.~\cite{Zhang14}.
Generally speaking, the inclusion of nuclear pasta does not lead to significant differences in
the thermodynamic quantities, but it may be important for the neutrino scattering rates
and elastic properties of stellar matter.

\section{Conclusions}
\label{sec:4}

In this work, we investigated the properties of nuclear pasta appearing in hot and dense
matter, associated with core-collapse supernovas and neutron-star mergers.
We employed the compressible liquid-drop (CLD) model to describe the pasta phases
with various geometric shapes. In the CLD model, the matter in the Wigner--Seitz cell is
assumed to separate into a dense liquid phase of nucleons and a dilute gas phase of
nucleons and $\alpha$ particles by a sharp interface.
The equilibrium conditions between the liquid and gas phases were derived by minimization
of the total free energy including the surface and Coulomb contributions, which are
clearly different from the Gibbs equilibrium conditions.
For the nuclear interaction, we employed the TM1e model with a small
symmetry energy slope $L=40$ MeV, which could be compatible with both
experimental nuclear data and recent observations of neutron stars.
To evaluate the influence of the density dependence of symmetry energy,
the results of the TM1e model were compared to those of the original TM1
model with a large symmetry energy slope $L=110.8$ MeV.
It is noteworthy that the TM1e and TM1 models have the same properties
of symmetric nuclear matter but different density dependencies of symmetry
energy, so that the comparison between the two models reflects the influence
solely from the symmetry energy without interference of the isoscalar part.

At given temperature $T$, proton fraction $Y_p$, and average baryon density $n_b$,
we performed calculations for all pasta phases considered, and then determined the
thermodynamically stable state with the lowest free energy.
The transition from nonuniform matter to uniform matter occurs at the density
where the free energy density of pasta phases becomes higher than that of homogeneous matter.
It was found that the inclusion of pasta phases could significantly delay the transition
to uniform matter as compared to the case with spherical nuclei only.
From the phase diagrams obtained, it was observed that at lower temperatures various
pasta shapes appear one by one with increasing density and their density ranges
are only weakly dependent on the temperature. At higher temperatures, the density
ranges shrink and some pasta shapes may have no chance to appear before the transition
to uniform matter. When the temperature reaches the critical value $T_c$,
nuclear pasta cannot be formed and the nonuniform matter phase disappears completely.
It was shown that the critical temperature $T_c$ depends on both the proton fraction $Y_p$
and the nuclear model used. Significant differences between the TM1e and TM1 models could
be observed in the phase diagram at the low-$Y_p$ region.
This implies that nuclear symmetry energy and its density dependence play a crucial role
in determining the properties of pasta phases in neutron-rich matter.

The present results with pasta phases using the CLD method were compared to those in
the realistic EOS table for astrophysical simulations, where the parametrized
Thomas--Fermi approximation was used and only spherical nuclei were taken into account.
It was found that thermodynamic quantities obtained in the two methods
are consistent with each other, but the inclusion of pasta phases causes small
discontinuities at the change of pasta shapes.
It is likely that the influence of pasta phases on the EOS for astrophysical
simulations is relatively limited. The discontinuities of the first-order phase
transition in pasta may play a role in the neutron-star cooling and affect the glitch phenomena.
Possible impacts of nuclear pasta on the neutrino scattering rates
need to be studied in future work.

\section*{Acknowledgment}

This work was supported in part by the National Natural Science Foundation of
China (Grants No. 11675083, No. 11775119, and No. 11805115).



\end{document}